\documentclass[prl,aps,twocolumn,floatfix,amsmath,amssymb,superscriptaddress]{revtex4-1}
\usepackage{latexsym}
\usepackage{graphicx}
\usepackage{tabularx}
\usepackage{times}
\usepackage[usenames,dvipsnames]{color}
\usepackage{amsmath}
\usepackage{dcolumn}
\usepackage{latexsym,amsmath,amssymb,bm,euscript}
\usepackage[
	colorlinks=true,
	urlcolor=BlueViolet,
 citecolor=Maroon,
	plainpages=false,
 	pdfpagelabels,
 	bookmarksnumbered
 	]{hyperref}

\renewcommand{\(}{\left(}
\renewcommand{\)}{\right)}

\definecolor{gray}{rgb}{0.4,.4,0.4}
\definecolor{purple}{rgb}{0.6,.0,0.6}
\definecolor{darkgreen}{rgb}{0.0,.6,0.}

\def\beq{\begin{equation}}
\def\eeq{\end{equation}}

\begin{document}

\title{Bosonic Analogue of Dirac Composite Fermi Liquid}

\author{David F. Mross}
\affiliation{Department of Physics and Institute for Quantum Information and Matter, California Institute of Technology, Pasadena, CA 91125, USA}

\author{Jason Alicea}
\affiliation{Department of Physics and Institute for Quantum Information and Matter, California Institute of Technology, Pasadena, CA 91125, USA}
\affiliation{Walter Burke Institute for Theoretical Physics, California Institute of Technology, Pasadena, CA 91125, USA}

\author{Olexei I. Motrunich}
\affiliation{Department of Physics and Institute for Quantum Information and Matter, California Institute of Technology, Pasadena, CA 91125, USA}
\affiliation{Walter Burke Institute for Theoretical Physics, California Institute of Technology, Pasadena, CA 91125, USA}

\begin{abstract}
We introduce a particle-hole-symmetric metallic state of bosons in a magnetic field at odd-integer filling. This state hosts composite fermions whose energy dispersion features a quadratic band touching and corresponding $2\pi$ Berry flux protected by particle-hole and discrete rotation symmetries. We also construct an alternative particle-hole symmetric state---distinct in the presence of inversion symmetry---without Berry flux. As in the Dirac composite Fermi liquid introduced by Son \cite{Son}, breaking particle-hole symmetry recovers the familiar Chern-Simons theory. We discuss realizations of this phase both in 2D and on bosonic topological insulator surfaces, as well as signatures in experiments and simulations.
\end{abstract}

\maketitle

{\bf \emph{Introduction.}}
The last year has seen numerous exciting developments in our understanding of electronic quantum-Hall states that resolved long-standing puzzles regarding particle-hole (PH) symmetry. At filling factor $\nu=\frac{1}{2}$, electrons fill exactly half of the available single-particle orbitals in the lowest Landau level (LLL). Within that subspace the system enjoys PH symmetry that is conspicuously absent in the classic Halperin-Lee-Read (HLR) theory \cite{HLR,WillettCFL,KangCFL,GoldmanCFL,DasSarmaBook,Willet,JainBook}. There, \emph{composite fermions}, obtained by attaching two fictitious flux quanta to electrons that cancel the applied field on average, fill a parabolic band and form a Fermi surface---i.e., a composite Fermi liquid (CFL). The corresponding Lagrangian density reads
\begin{align}
{\cal L}_\text{CS} =& f^\dagger\left[i\( D_0- i A_0\)-\frac{(\vec D- i \vec A)^2}{2 m^*}\right]f- \frac{k}{8\pi} \epsilon_{\kappa\mu\nu}a_\kappa \partial_\mu a_\nu, \label{eqn.hlr}
\end{align}
where $f$ is the composite fermion field, $A_\mu$ (with $\mu=0,1,2$) is the electromagnetic vector potential, $D_\mu = \nabla_\mu - i a_\mu$ denotes the covariant derivative with $a_\mu $ an emergent gauge field, and $k=1$ is the level of the Chern-Simons term that attaches flux. Despite the absence of PH symmetry, HLR theory is remarkably successful in predicting experimental results at and around $\nu=\frac{1}{2}$.

To incorporate PH symmetry, Son proposed that composite fermions are Dirac particles at finite density coupled to an emergent gauge field \cite{Son} without a Chern-Simons term:
\begin{align}
{\cal L}_\text{QED$_3$} = i \bar\Psi D_\mu \gamma^\mu \Psi-\frac{k}{4\pi}\epsilon_{\kappa\mu\nu}A_\kappa \partial_\mu a_\nu.\label{eqn.son}
\end{align}
Here $\Psi$ and $\bar \Psi = \Psi^\dagger \gamma^0$ are two-component spinors while $\gamma^\mu$ are Dirac matrices. Equation~\eqref{eqn.son} implements two important features of the half-filled Landau level: $(i)$ the Dirac composite fermions are neutral under the external vector potential $A_\mu$ \cite{Read94,ShankarMurthy97,DHLee98,Pasquier98} and $(ii)$ the theory preserves the anti-unitary PH transformation
\begin{align}
&{\cal C} \Psi {\cal C }^{-1}= i \sigma^2 \Psi\ ,\ \ \ \ &{\cal C} (a_0,\vec a){\cal C }^{-1}=(a_0,-\vec a).
\end{align}
Several subsequent works support Son's theory and the presence of PH symmetry at $\nu = \frac{1}{2}$ \cite{MetlitskiVishwanath2015,WangSenthilReview,Geraedts,ShankarMurthy,Kachru,diracduality,MulliganPH,BalramJain}.

These developments prompt us to revisit the CFL formed by bosons at filling factor $\nu=1$, where flux attachment yields Eq.~\eqref{eqn.hlr} with $k=2$. An important conceptual difference between bosonic quantum-Hall states and their fermionic counterparts is the absence of a single-particle description of integer quantum-Hall (IQH) states, which in turn obscures a precise definition of PH symmetry even when restricting to the LLL. To access PH-symmetric CFLs of $\nu = 1$ bosons, we therefore follow a two-pronged approach: First, we study bosons at a `plateau transition' \footnote{Technically, the model studied here requires disorder or a superlattice to describe a plateau transition.} between a $\nu=2$ IQH state \cite{senthillevin,LuVishwanath,Geraedts2013288} and the vacuum. Upon fine-tuning, the critical theory exhibits a PH symmetry analogous to the electronic case, in addition to microscopic inversion symmetry. Second, we consider the surface of a particular 3D symmetry protected topological phase (SPT) of bosons \cite{AshvinSenthil}, where both symmetries can be realized microscopically. 

\begin{figure*}
 \includegraphics[width=.90\textwidth]{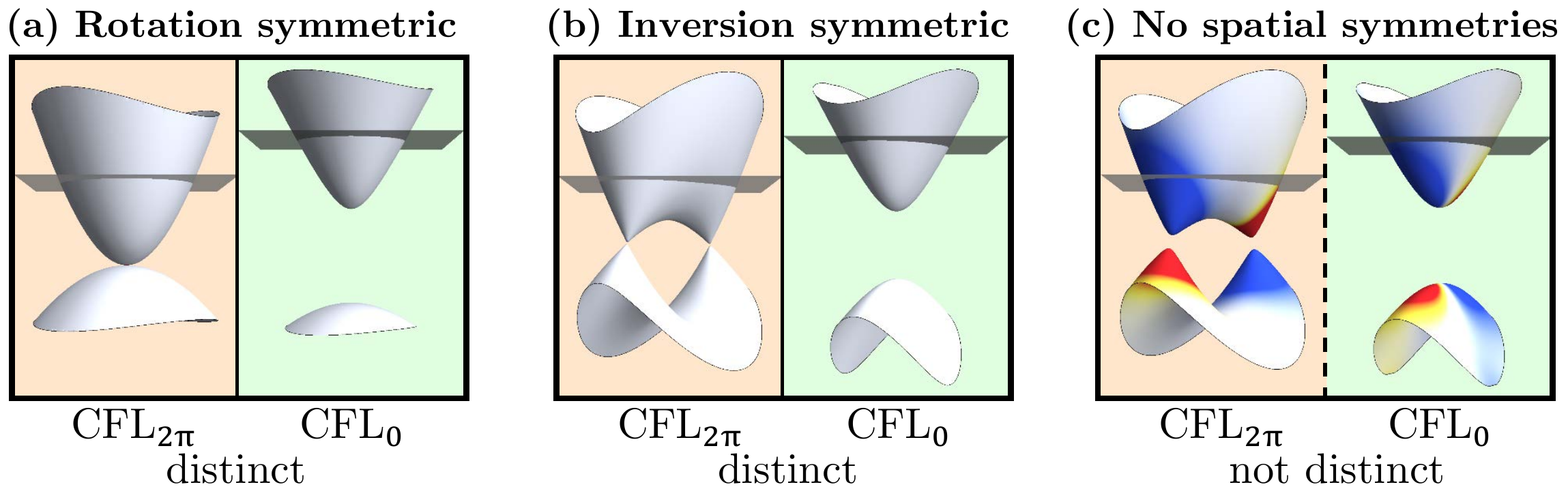}
 \caption{Composite-fermion band structures for CFL$_{2\pi}$ and CFL$_0$ states with particle-hole symmetry and varying spatial symmetries. Horizontal planes indicate the chemical potentials of interest. (a) With four-fold rotation symmetry CFL$_{2\pi}$ exhibits a protected quadratic band touching, in sharp contrast to the gapped spectrum of CFL$_0$. (b) Breaking rotation down to inversion symmetry allows the band touching to split into two protected Dirac cones but preserves the $2\pi$ Berry flux. (c) Lifting inversion symmetry generically produces a gapped spectrum with non-zero Berry curvature indicated by color (blue and red for opposite sign). The sharp distinction between CFL$_{2\pi}$ and CFL$_{0}$ then disappears. Away from degeneracy points, particle-hole and inversion symmetries respectively constrain the Berry curvature ${\cal B}$ at momentum ${\bf k}$ as ${\cal B}({\bf k})=-{\cal B}(-{\bf k})$ and ${\cal B}({\bf k})={\cal B}(-{\bf k})$; the presence of both symmetries [i.e., cases (a) and (b)] thus implies zero Berry curvature.}\label{fig.quad}
\end{figure*}

These methods suggest a natural bosonic analogue of Eq.~\eqref{eqn.son} given by
\begin{align}
{\cal L}_\text{CFL$_{2\pi}$} =\ & \Psi^\dagger \begin{pmatrix} iD_0 & (D_1 + i D_2)^2\\ (D_1- i D_2)^2& iD_0 \end{pmatrix} \Psi\nonumber\\
&-\frac{1}{2\pi} \epsilon_{\kappa\mu\nu}A_\kappa \partial_\mu a_\nu \ .\label{eqn.bosonlagrangian}
\end{align}
Note that the composite-fermion density is dynamically fixed to $n_\text{CF}=(\partial_1 A_2 - \partial_2 A_1)/2\pi$.
The first line of Eq.~\eqref{eqn.bosonlagrangian} also describes electronic excitations of bilayer graphene near one of the two valleys \cite{grapheneRMP}. In the present context, the composite fermions analogously exhibit a single quadratic band touching that is protected against weak perturbations that respect both PH symmetry ${\cal C}$ and fourfold rotation symmetry ${\cal R}(\pi/2)$ (see Fig.~\ref{fig.quad}). When only inversion symmetry ${\cal I}$ is present, the spectrum remains gapless but the band touching generically splits into two Dirac cones, similar to the effect of trigonal warping in bilayer graphene \cite{grapheneRMP}. At suitable doping, either case features a single Fermi surface enclosing $2\pi$ Berry flux. To distinguish different kinds of PH-symmetric CFLs we adopt notation CFL$_{k\pi}$, where $k\pi$ is the Berry flux enclosed in the composite-fermion Fermi surface. Thus, the fermionic $\nu=\frac{1}{2}$ state described by ${\cal L}_\text{QED$_3$}$ corresponds to CFL$_{\pi}$ while the bosonic state described by ${\cal L}_\text{CFL$_{2\pi}$}$ is CFL$_{2\pi}$. As we will see, an alternative ${\cal C}$- and ${\cal I}$-symmetric state for bosons, CFL$_{0}$, with zero Berry flux is also possible. For any CFL$_{k\pi}$, weak PH symmetry breaking yields a gapped spectrum, and subsequently integrating out the negative energy states generates a level-$k$ Chern-Simons term as in ${\cal L}_\text{CS}$ (i.e., no Chern-Simons term for $k=0$). Absent inversion symmetry, the sharp distinction between CFL$_{2\pi}$ and CFL$_{0}$ disappears.

\begin{figure}[h]
 \includegraphics[width=.9\columnwidth]{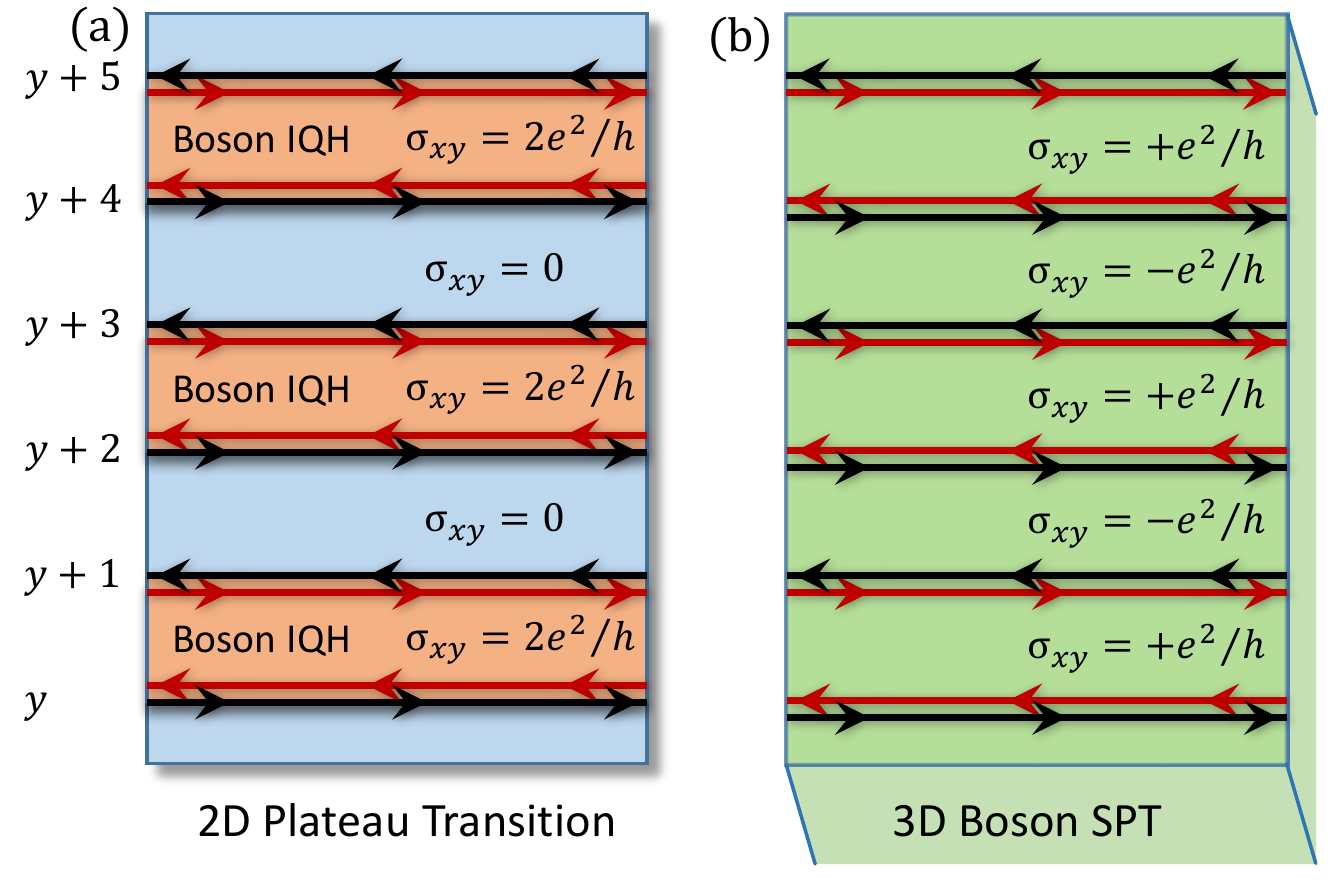}
 \caption{(a) Alternating strips of bosons at $\nu=2$ and $\nu=0$ correspond to an average filling factor $\nu=1$. Edges contain a chiral charge mode and a counter-propagating neutral mode, described by the K-matrix $\sigma^x$. (b) The same edge-state network arises on the surface of a 3D bosonic SPT with $U(1)\times{\cal C}$ symmetry when the antiunitary PH-symmetry ${\cal C}$ is broken oppositely for neighboring strips.}\label{fig.strip}
\end{figure}

{\bf \emph{Bosons at $\nu=1$ as a plateau transition.}} Consider a system composed of narrow strips of width $d$ along the $y$ direction and infinite along $x$; see Fig.~\ref{fig.strip}(a). The boson density $\rho$ for adjacent strips alternates between $\rho_0$ and $0$, and a uniform perpendicular magnetic field $B = \frac{c h \rho_0}{2e}$ yields filling factor $\nu=\frac{c h \rho}{eB}=2$ for the $\rho_0$ strips. At length scales much larger than $d$ we thus obtain bosons with average filling $\nu=1$. 

We require the $\nu=2$ strips to form bosonic IQH states \cite{senthillevin,LuVishwanath,Geraedts2013288}. It is useful to view a given $\nu = 2$ strip as composed of quantum wires \cite{TeoKaneChains} labeled by $j=1,\ldots, N$ (not to be confused with the domain-wall labels $y$ in Fig.~\ref{fig.strip}). Each wire hosts charge-$e$ bosons $ \sim e^{i \varphi_j}$ described by 
\begin{align}
 {\cal L }_\text{wire} =\frac{\partial_x \theta_j}{\pi}(\partial_t \varphi_j-A_{0})+ \frac{v}{2\pi}[(\partial_x \varphi_j-A_{1})^2+(\partial_x \theta_j)^2]\nonumber
\end{align}
where $\frac{\partial_x \theta_j}{\pi}$ is conjugate to $\varphi_j$. When a boson hops between neighboring wires at non-zero magnetic field---conveniently taken in the gauge $A_2 = B x$---it acquires an Aharonov-Bohm phase $e^{ i\frac{2\pi e d B}{h c} x}$ that prevents condensate formation. These oscillating phases can be compensated, however, when a phase slip $\sim e^{2 i \theta+ i 2 \pi d \rho_0 x}$ accompanies boson hopping. For $\nu=2$ this occurs for second-neighbor hopping described by
\begin{align}
{\cal L}_\text{boson IQH}= g_\text{IQH}\sum\nolimits_j \cos\(\varphi_{j+1} - \varphi_{j-1} + 2\theta_{j}\)\label{eq.bosonint}.
\end{align}
A bosonic IQH state with $\sigma_{xy} = 2e^2/h$ \cite{senthillevin,LuVishwanath,Geraedts2013288} emerges when $g_\text{IQH}$ flows to strong coupling. The $\nu=2$ strip then hosts edge states with two flavors $\alpha=\pm$ of charge-$e$ bosons $b_{y = 1(2),\alpha}\sim e^{i\phi_{y = 1(2),\alpha}}$ at the lower (upper) edge, where
\begin{align}
 &\phi_{y = 1,+} \equiv \varphi_1, \ \ \ \ \ \ \ \ \ \ \ \ \ \ \ \ \ \ \ \ \ \ \phi_{y = 1,-} \equiv \varphi_2+2 \theta_1, \nonumber \\
 &\phi_{y = 2,+} \equiv \varphi_{N-1}-2 \theta_{N} ,\ \ \ \ \ \phi_{y = 2,-}\equiv \varphi_{N} .
 \label{EdgeFields}
\end{align}
The Lagrangian density for the lower and upper edges is succinctly written as
\begin{align}
{\cal L }_\text{edge} &= \frac{(-1)^yK_{\alpha\alpha'}}{4\pi}\partial_x \phi_{y,\alpha}\partial_t\phi_{y,\alpha'}+ \frac{u}{4\pi}\(\partial_x \phi_{y,\alpha}\)^2.\label{eqn.sigmaxstate}
\end{align}
Here $K=\sigma^x$ and $\alpha,\alpha'$ are implicitly summed here and below. Equation~\eqref{eqn.sigmaxstate} generalizes straightforwardly to the full 2D system in Fig.~\ref{fig.strip}(a) when edges are enumerated by integers $y$. 

To access 2D CFL's we allow tunneling between neighboring edges. For example, flavor-conserving tunnelings read
\begin{align}
 {\cal L}_\text{hop}= w_{y,\alpha}\left( e^{i\frac{2h e B}{h c}d N x} b^\dagger_{y+1,\alpha}b_{y,\alpha}+c.c.\right),
\end{align} 
i.e., the edge bosons $b_{y,\alpha}$ experience a uniform magnetic field. Our setup preserves a microscopic inversion symmetry,
\begin{align}
 {\cal I}b_{y,\alpha}(x){\cal I}^{-1}=b_{1-y,-\alpha}(-x)\ \ \ \ \ \text{(inversion)},
 \label{inversion}
\end{align}
[see Eqs.~\eqref{EdgeFields}] that constrains the hopping amplitudes via $w_{y,\alpha} = w_{-y,-\alpha}$. In the case of translation invariance $y\rightarrow y +2$, only $w_{y=\text{even}}$ (hopping across vacuum) and $w_{y=\text{odd}}$ (hopping across IQH strips) are independent.
When these are fine-tuned to be equal, the low energy theory ${\cal L}_\text{hop}+{\cal L }_\text{edge} $ additionally exhibits an emergent anti-unitary PH symmetry,
\begin{align}
 {\cal C}b_{y,\alpha}{\cal C}^{-1}=b_{y+1,\alpha}^\dagger \ \ \ \ \ \text{(particle-hole)}. 
 \label{eqn.bosonph}
\end{align} 

{\bf \emph{3D boson SPT surface.}}
Closely related physics can appear at the surface of a 3D bosonic SPT \cite{VishwanathSenthil2013} with a conserved charge that is odd under a local anti-unitary PH symmetry, i.e., $U(1)\times {\cal C}_{\rm local}$ \cite{WangSenthilBosons} \footnote{One can alternatively view PH as a time-reversal symmetry, as done earlier in Ref.~\onlinecite{VishwanathSenthil2013} where this SPT was labeled $U(1)\times Z_2^\text{T}$.}. This symmetry pins the chemical potential associated with the conserved charge to zero but permits an orbital magnetic field. Breaking ${\cal C}_{\rm local}$ generates a gapped, unfractionalized surface with Hall conductance $\sigma_{xy}=\pm e^2/h$. At boundaries between domains with oppositely broken ${\cal C}_{\rm local}$, the Hall conductance changes by $2e^2/h$---implying the existence of gapless edge states described by Eq.~\eqref{eqn.sigmaxstate}. Consider now a situation where the SPT surface hosts alternating $\pm e^2/h$ strips of equal width [Fig.~\ref{fig.strip}(b)]. Such a surface breaks ${\cal C}_{\rm local}$ but retains this symmetry when composed with a translation $T_y$ by one strip width, i.e., Eq.~\eqref{eqn.bosonph} with ${\cal C} = {\cal C}_{\rm local}T_y$. Note that unlike the 2D plateau transition, ${\cal C}$ is a true microscopic symmetry on the SPT surface. 

{\bf \emph{Bosonic CFL's.}}
Both the plateau transition and the SPT surface realization lead us to study the theory ${\cal L }_\text{edge} + {\cal L}_\text{hop}$. Our analysis is facilitated by an explicit duality mapping for network models of this kind \cite{diracduality} relating the surface states of 3D topological insulators to quantum electrodynamics in $(2+1)$ dimensions (QED$_3$) described by ${\cal L}_\text{QED$_3$}$ (see also Refs.~\onlinecite{MetlitskiVishwanath2015,WangSenthil2015,wangsenthilu1,XuYouSPT}). The density of dual composite fermions is proportional to the physical magnetic field $B$, while the number of flavors $N_f$ depends on the statistics of the microscopic particles forming the 3D TI:
\begin{align}
&\text{3D electron TI surface} \ \ \ \leftrightarrow \ \ \ \text{$N_f=1$ QED}_3,\nonumber\\
&\text{3D boson TI surface} \ \ \ \ \! \ \ \ \leftrightarrow \ \ \ \text{$N_f=2$ QED}_3\nonumber.
\end{align} 
Either bosonic setup from Fig.~\ref{fig.strip} thus maps to ${\cal L}_\text{QED$_3$}$ in Eq.~\eqref{eqn.son} with $k = 2$ and $N_f=2$ fermion flavors.

We will package the two flavors into a single four-component spinor $\Psi_4$ and use Dirac matrices $\gamma^0 = \tau^0\sigma^z$, $\gamma^1=i \tau^0\sigma^y$, $\gamma^2 =- i \tau^0\sigma^x$, where $\sigma^\mu$ and $\tau^\mu$ are respectively intra- and inter-flavor Pauli matrices. Following the mapping from Ref.~\onlinecite{diracduality}, the symmetries in Eqs.~\eqref{inversion} and \eqref{eqn.bosonph} act as \footnote{We performed a change of basis relative to Ref. \onlinecite{diracduality} that results in $\tau^{z}\sigma^{z}\rightarrow \tau^{x}\sigma^{x}$.}
\begin{align}
{\cal C} \Psi _4{\cal C}^{-1} &= \sigma^y \tau^y \Psi _4 ,\ \ \ \ \ \ \ \ \ \ \ \ \
{\cal I} \Psi _4{\cal I}^{-1} = \sigma^z \tau^z \Psi _4 .
\end{align}
The continuum dual QED$_3$ theory also preserves continuous rotations
\begin{align}
{\cal R}(\Phi)\Psi _4(\vec x){\cal R}^{-1}(\Phi)=e^{i\frac{ \Phi}{2}(\tau^z + \sigma^z)}\Psi _4\left[{\cal R}(\Phi)\vec x\right]
\end{align}
with ${\cal R}(\pi)={\cal I}$. While ${\cal R}$ is not a microscopic symmetry of the network-model, we expect it to be relevant for CFL realizations in isotropic systems.
\begin{figure}
 \includegraphics[width=\columnwidth]{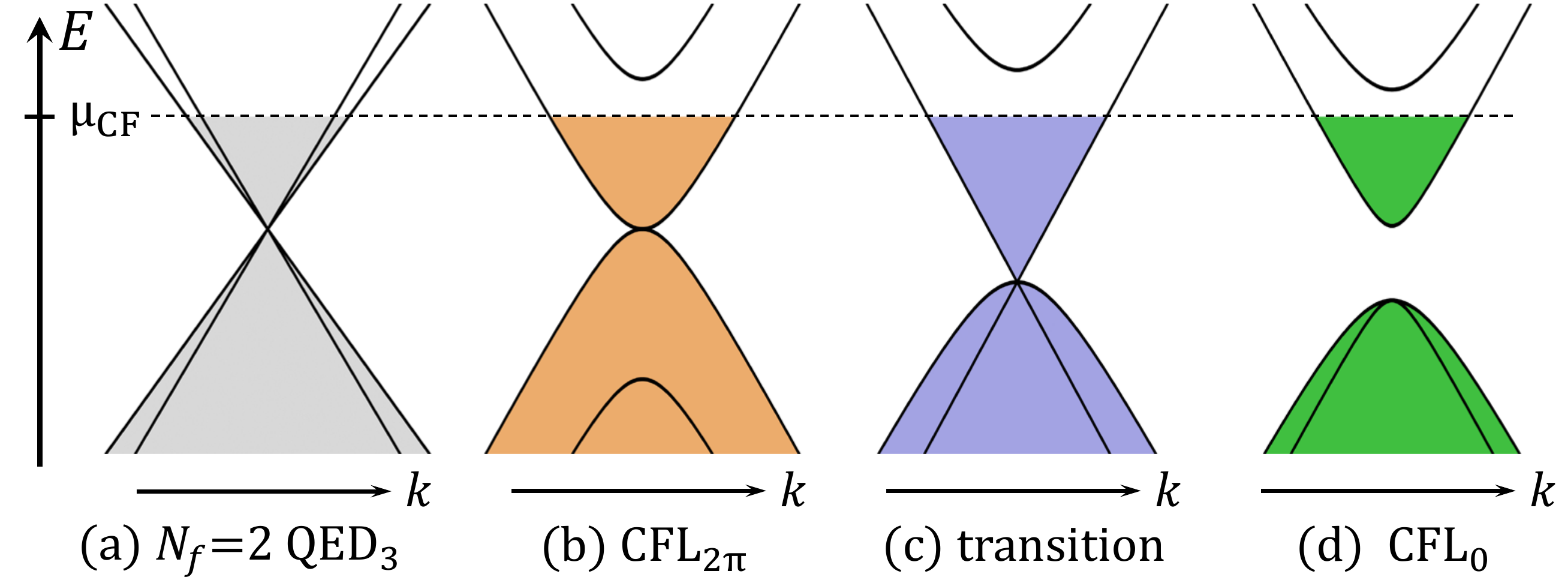}
 \caption{ (a) Massless Dirac band structure for `pure' $N_f = 2$ QED$_3$ that is dual to the bosonic models studied here (shown with different velocities to emphasize $N_f=2$). Magnetic field for the bosons maps to a non-zero composite fermion density that yields two Fermi surfaces. (b) PH and rotation-symmetric perturbations with $|g|>|\Delta|$ yield the CFL$_{2\pi}$ with a quadratic band touching and a single Fermi surface. (c) The transition between CFL$_{2\pi}$ and CFL$_{0}$ occurs at $|g|=|\Delta|$ where three bands meet at $k=0$. (d) For $|g|<|\Delta|$ one obtains CFL$_{0}$ featuring a gapped spectrum and a single Fermi surface at appropriate density. Note that the nature of the partially filled positive-energy band changes qualitatively, with Berry flux jumping from $2\pi$ to $0$. }
 \label{fig.2dbands}
\end{figure}

We thus first analyze a composite-fermion Hamiltonian $ H_{{\cal R}}=\Psi_4^\dagger h_{{\cal R}}\Psi_4$ containing general momentum-independent bilinears preserving both ${\cal C}$ and ${\cal R}(\pi/2)$ \footnote{An additional allowed term $g'(\tau^x \sigma^y - \tau^y \sigma^x)$ can be rotated onto $g$, see Appendix.},
\begin{align}
 h_{{\cal R}} = i\sigma^x D_1 +i\sigma^y D_2 -\Delta \tau^z \sigma^z+g(\tau^x \sigma^x + \tau^y \sigma^y).
\end{align}
At mean-field level (neglecting $a_\mu$) the spectrum contains four bands which we label as positive and negative according to their large-$k$ asymptotics, i.e.,
\begin{align}
&E_\text{positive}( \vec k) = \pm g+ \sqrt{ \vec k^2 + (g\pm \Delta)^2}\label{eq.spec1},\\
&E_\text{negative}( \vec k) = \pm g-\sqrt{ \vec k^2 + (g\pm \Delta)^2} \label{eq.spec2}.
\end{align}
Several distinct regimes are accessible depending on $g,\Delta$ as sketched in Figs.~\ref{fig.2dbands}(a)-(d):

(a) For $g = \Delta = 0$ we recover two massless Dirac cones.

(b) At finite $|g|>|\Delta|$ a quadratic band-touching emerges;
the CFL$_{2\pi}$ state then appears when the chemical potential intersects only one of the central bands as in Fig.~\ref{fig.2dbands}(b). In this case one can project onto states close to the band touching (see Appendix), yielding Eq.~\eqref{eqn.bosonlagrangian} with two-component spinors $\Psi_2$ that transform as
\begin{align}
{\cal C}\Psi_2 {\cal C}^{-1}=\sigma^x \Psi_2,\ \ \ \ \ \ {\cal R}(\Phi)\Psi_2 {\cal R}^{-1}(\Phi)= e^{ i \Phi \sigma^z}\Psi_2.\label{psi2transform}
\end{align}
The only perturbation allowed by ${\cal R}(\pi/2)$ up to ${\cal O}(k^2)$ is the mass term $\Psi_2^\dagger\sigma^z\Psi_2$---which is odd under ${\cal C}$. Thus the CFL$_{2\pi}$ with quadratic band touching is stable with these symmetries. Relaxing ${\cal R}(\pi/2)\rightarrow {\cal R}(\pi)={\cal I}$ allows the terms $\Psi_2^\dagger\sigma^{x,y}\Psi_2$, which split the band touching into two Dirac cones without opening a gap [see Fig.~\ref{fig.quad}(b)]. Upon breaking inversion, the PH-symmetric terms $\Psi_2^\dagger\sigma^{z}i \partial_{1,2}\Psi_2$ gap out the two Dirac cones [Fig.~\ref{fig.quad}(c)]. 

(c) At $|g| = |\Delta|$ the spectrum hosts a three-fold band touching. 

(d) For $|g| < |\Delta|$ a gap opens and the conduction band `detaches' from the valence bands (see Appendix for details). The special point (c) thus marks the transition at which the topological winding associated with the quadratic band touching transfers to the bottommost bands (for $\Delta > 0$; with $\Delta<0$ the band order reverses). Integrating out these filled negative-energy bands does not generate a Chern-Simons term. At suitable doping one thus obtains a single Fermi surface with neither a Chern-Simons term nor Berry curvature, corresponding to CFL$_{0}$. 

\begin{figure}
 \includegraphics[width=\columnwidth]{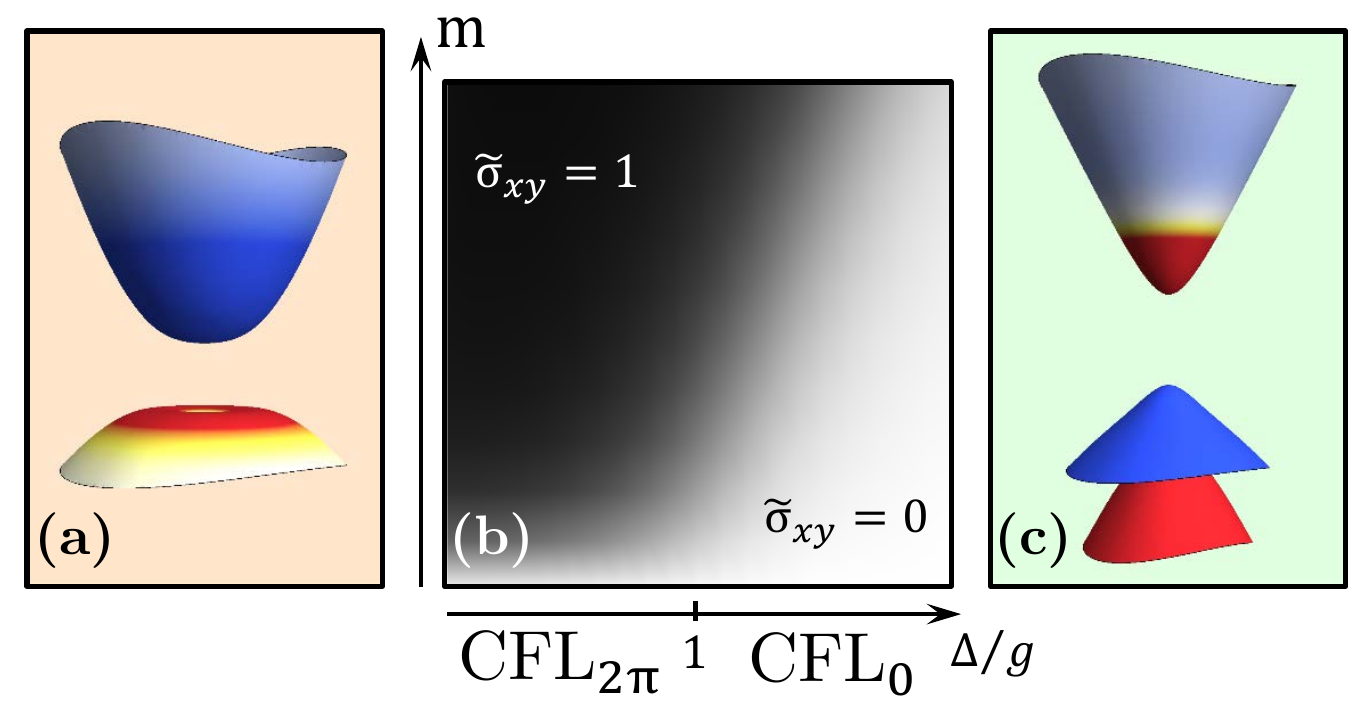}
 \caption{Both CFL$_{2\pi}$ and CFL$_{0}$ exhibit vanishing composite-fermion Hall conductance $\tilde \sigma_{xy}=0$ due to PH symmetry ${\cal C}$. Upon breaking ${\cal C}$ by the term $m \Psi_2^\dagger\sigma^z\Psi_2$, $\tilde \sigma_{xy}=0$ crosses over to the HLR value $\tilde \sigma_{xy}=1$ (center). For $\Delta/g<1$ the CFL$_{2\pi}$ band touching is lifted by $g$, resulting in bands with opposite Berry-curvature (left). For $ \Delta/g >1$ the partially filled CFL$_{0}$ band develops Berry curvature as a function of $m$ and changes sign between small and large momenta (right).}
 \label{fig.hall}
\end{figure}

We emphasize that the distinction between CFL$_{2\pi}$ and CFL$_{0}$ requires \emph{both} PH and inversion symmetries, since breaking either generically produces a gapped spectrum. When PH is broken, generic Fermi surfaces enclose a non-universal non-quantized Berry flux (Fig.~\ref{fig.hall}). On the other hand, breaking inversion symmetry while preserving PH always yields zero enclosed Berry flux [Fig.~\ref{fig.quad}(c)].
 
{\bf \emph{Properties.}} It is instructive to analyze how both CFL$_{2\pi}$ and CFL$_0$ reduce to the conventional HLR state upon breaking PH symmetry. A useful quantity in this context is the composite-fermion Hall conductance $\tilde \sigma_{xy} = \frac{k}{2}-\frac{\gamma}{2\pi}$, where $k \in \mathbb{Z}$ is the level of the Chern-Simons term for the emergent gauge field and $\gamma \in \mathbb{R}$ is the Berry flux enclosed in the Fermi surface that yields an anomalous Hall effect \cite{Son}. In the bosonic HLR state [Eq.~\eqref{eqn.hlr}] $k=2$ and $\gamma=0$ so that $\tilde \sigma_{xy}=1$. PH symmetry, however, demands $\tilde \sigma_{xy}=0$ in both CFL$_{2\pi}$ and CFL$_0$. For the former, breaking PH symmetry via $m \Psi_2^\dagger\sigma^z\Psi_2$ splits the bands [Fig.~\ref{fig.hall}(a)], whereupon integrating out the negative energy states generates a $k=2$ Chern-Simons term. In contrast, weakly breaking PH symmetry in CFL$_{0}$ does not produce a Chern-Simons term but induces non-zero Berry curvature in the partially filled band; see Fig.~\ref{fig.hall}(c) and the Appendix for details. The two cases can be summarized as:

\begin{align}
&k_\text{CFL$_{2\pi}$}=2,\ \  \ \gamma_\text{CFL$_{2\pi}$}=2\pi \left[1- \frac{2(g-\Delta)m}{\sqrt{\(2(g-\Delta)m\)^2+K_F^4}}\right],\nonumber\\
&k_\text{CFL$_{0}$}=0,\ \ \ \  \ \gamma_\text{CFL$_{0}$}=-2\pi \frac{m K_F^2}{8 g \Delta^2}+{\cal O} \(\frac{K_F^4}{\Delta^4},\frac{m^2}{g^2}\),\label{eqnberryphases}
\end{align}
where $K_F$ is the Fermi wavevector. 
As Fig.~\ref{fig.hall}(b) illustrates, when $m \rightarrow \infty$ we asymptotically recover $\tilde \sigma_{xy} = 1$ starting from CFL$_{2\pi}$ or CFL$_0$, consistent with the HLR result. While $\tilde \sigma_{xy}$ is not directly observable, Ref.~\onlinecite{Son} suggested that it may be determined from detailed Hall-effect measurements; Ref.~\onlinecite{PotterCFL} suggested the Nernst effect as a more sensitive $\tilde \sigma_{xy}$ probe. 

A useful device for determining the presence of PH symmetry in numerical studies is via $2K_F$ oscillations in the composite-fermion density $\tilde \rho$. In the electronic case, $\tilde \rho_{k \approx 2 K_F}$ is PH-odd and thus generically contributes to the physical charge density \cite{Geraedts}. In the bosonic CFL$_{2\pi}$ and CFL$_0$, by contrast, $\tilde \rho_{k \approx 2 K_F}$ is PH-even and does not contribute to the boson density $\rho$. Any $2K_F$ oscillations in the boson density thus directly probe PH-symmetry breaking.

The distinction between CFL$_{2\pi}$ and CFL$_{0}$ in the presence of both ${\cal I}$ and ${\cal C}$ is more subtle; operators constructed from composite fermions near the Fermi surface do not distinguish between the two. Still, the two clearly differ in the limit of low composite-fermion density $K_F \ll \Delta,g$ in the same way that bilayer graphene is distinct from 2D electron gases with parabolic dispersion \cite{novoselovblg} (see also the Appendix).

{\bf \emph{Gapped phases.}}
It is well known that Cooper-pairing composite fermions generates a quantum-Hall insulator of the microscopic particles. As with the HLR theory, it is natural in CFL$_{2\pi}$ or CFL$_{0}$ to consider chiral, odd-angular-momentum pairing between composite fermions---which permits a full gap for spinless fermions. Such states always (spontaneously) break ${\cal C}$ \cite{WangSenthilBosons}. An alternative gapped phase arose in the study of time-reversal-symmetric surfaces of 3D bosonic SPTs in Ref.~\onlinecite{VishwanathSenthil2013}. In our network-model this phase arises from the edge-boson interaction
\begin{align}
{\cal L}_{2\sigma^x} &=\sum\nolimits_{\alpha,y} u_\alpha\cos \(\phi_{y-1,\alpha} - 2 \phi_{y,\alpha} + \phi_{y+1,\alpha}\),
\end{align}
where $u_+ \neq u_-$ results in ${\cal C}$-symmetric---but not ${\cal I}$-symmetric---topological order with $K=2\sigma^x$ \cite{STO}. (${\cal I}$ interchanges $u_+$ and $u_-$, with the gap closing when $u_+ = u_-$.) 

A fully symmetric gapped state is nevertheless readily constructed as a composite-fermion superconductor driven by ${\cal L}_\text{Pair}\sim \Psi_{4}^\dagger \sigma^y\tau^x\Psi_{4}^\dagger+\text{H.c.}$. We expect that this state corresponds to a `larger' topological order that can be reduced to $K=2 \sigma^x$ by condensing an ${\cal I}$-odd boson. Edge-boson interactions that generate such a state may be obtained following Refs.~\onlinecite{STO,diracduality,WTI}, but that is not our focus. We simply note that in CFL$_{2\pi}$ and CFL$_{0}$ with a single Fermi surface, ${\cal L}_{2\sigma^x}$ and ${\cal L}_\text{Pair}$ are both absent in the projected Hilbert space. Accessing either ${\cal C}$-symmetric gapped state requires a finite coupling strength. Conversely, any gapped state emerging from a weak-coupling instability of CFL$_{2\pi}$ or CFL$_{0}$ necessarily breaks ${\cal C}$.

{\bf \emph{Conclusions.}}
We constructed two PH-symmetric metallic states, dubbed CFL$_{2\pi}$ and CFL$_{0}$, for bosons at $\nu=1$. These phases are distinct provided PH and inversion symmetries are present. In either case, $2K_F$ oscillations in the physical boson density are absent but appear when PH symmetry is broken. Furthermore, once PH symmetry is (weakly) broken the crossover between these states and conventional HLR theory may be observed in transport measurements. We also elucidated the relationship between PH-symmetric CFLs and gapped quantum Hall states, such as the bosonic Moore-Read state which breaks PH symmetry, and the $K = 2 \sigma^x$ state which does not.

A recent study by Wang and Senthil \cite{WangSenthilBosons} considers bosons at $\nu=1$ in the LLL with PH symmetry and proposes 
a CFL with Berry phase $-2\pi$. We believe that the states introduced here are closely related; we emphasize however that inversion symmetry is crucial in our setup to define a Berry phase of $2\pi$.

{\bf \emph{Acknowledgments.}} We gratefully acknowledge T.~Senthil, C.~Kane, M.~Metlitski, and D.~T.~Son for valuable discussions. This work was supported by the NSF through grants DMR-1341822 (JA) and DMR-1206096 (OIM); the Caltech Institute for Quantum Information and Matter, an NSF Physics Frontiers Center with support of the Gordon and Betty Moore Foundation; and the Walter Burke Institute for Theoretical Physics at Caltech.

\bibliography{bosonqh}
\newpage
\appendix
\section{\large General composite-fermion band structure}
In the absence of any symmetry, the most general momentum-independent bilinear perturbations to the composite-fermion Hamiltonian 
\begin{align}H_0 = \Psi_4^\dagger (\sigma^x k_1 + \sigma^y k_2)\Psi_4\end{align}
are given by 
\begin{align}
&\delta H=\sum_{\alpha,\beta=0,x,y,z} c_{\alpha \beta}\hat h_{\alpha\beta},\\
&\hat h_{\alpha\beta}\equiv\Psi_4^\dagger \sigma^\alpha \tau^\beta\Psi_4,\label{eqn.halphabeta}
\end{align} 
where $c_{\alpha\beta}$ are real constants. We now discuss these terms according to their symmetries:
\begin{enumerate}

 \item {\bf Particle-hole symmetry broken.} The six terms $\hat h_{i0}$ and $\hat h_{0i}$ with $i=x,y,z$, are odd under particle-hole symmetry. Among these, $\hat h_{x/y,0}$ and $\hat h_{0,x/y}$ break rotation symmetry; the former moves the Dirac cones in momentum while the latter splits them in energy. The terms $\hat h_{z0}$ and $\hat h_{0z}$ are rotation symmetric; the former opens a gap while the latter is a chemical potential with opposite sign for the two cones.

 \item {\bf Particle-hole and rotation symmetries.} The combination of particle-hole and (fourfold) rotation symmetry allows only the terms $\hat h_{zz}$, $\hat h_{xx}+\hat h_{yy}$, and $\hat h_{xy}-\hat h_{yx}$. The latter two can be turned into each other using a unitary transformation $e^{i \theta \tau^z}$ which commutes with ${\cal C}$ and ${\cal R}(\Phi)$. One may therfore without loss of generality restrict the analysis to $\hat h_{zz}$ and $\hat h_{xx}+\hat h_{yy}$ only, as we did in the main text. Depending on the relative magnitude of $\hat h_{zz}$ and $\hat h_{xx}+\hat h_{yy}$ one either finds a quadratic band touching (CFL$_{2\pi}$), or a gapped spectrum (CFL$_{0}$).
 
 \item {\bf Particle-hole and inversion symmetries.} When the rotation symmetry is broken down to twofold rotations (i.e., inversion), additional terms $\hat h_{xx}-\hat h_{yy}$ and $\hat h_{xy}+\hat h_{yx}$ are allowed. In the CFL$_{2\pi}$ regime, their effect is to split the quadratic band touching into two Dirac cones at different momenta. Inversion and particle-hole symmetries map these cones onto one another and protect them from opening a gap.
 
 \item {\bf Particle-hole symmetry only.} When particle-hole is the only symmetry present, $\hat h_{zx}$, $\hat h_{zy}$, $\hat h_{xz}$ and $\hat h_{yz}$ are also allowed and generically give rise to a gapped spectrum.

\end{enumerate}

\section{\large Diagonalization of ${\cal C}$- and ${\cal R}(\pi/2)$-symmetric Hamiltonian}
We consider the Hamiltonian $H_{\cal R} = \Psi^\dagger_4 h_{\cal R} \Psi_4$ with
\begin{align}
h_{\cal R}(\vec k) = 
\begin{pmatrix}
-\Delta & 	k e^{- i \phi_k} &	 0 & 	0 \\
k e^{ i \phi_k} & \Delta&	2g& 	0 \\
0 & 	2g &	 \Delta & 	k e^{- i \phi_k} \\
0 & 	0&	k e^{ i \phi_k} & -\Delta\label{eqn.rotandph}
\end{pmatrix},
\end{align}
where $ke^{i \phi_k}= k_1 + i k_2$. The eigenvalues of $h_{\cal R}(\vec k)$ are
\begin{align}
E_1( k) &= \sqrt{ k^2+g_+^2}+g \nonumber, \\
E_2( k) &= \sqrt{ k^2+g_-^2}-g \nonumber ,\\
E_3( k) &= -\sqrt{ k^2+g_+^2}+g \nonumber, \\
E_4( k) &= -\sqrt{ k^2+g_-^2}-g \nonumber ,
\end{align}
where $g_\pm \equiv g \pm \Delta$ and $E_1>E_2>E_3>E_4$ for $g>0$. The corresponding normalized eigenvectors are given by
\begin{align}
&u_{1/3}(\vec k) =\frac{1}{\sqrt{2 +2 \frac{\(g_+ \pm \sqrt{k^2+g_+^2}\)^2}{k^2}}} \begin{pmatrix}
 e^{-i \phi_k }\\ \frac{g_+ \pm \sqrt{k^2+g_+^2}}{k}\\ \frac{g_+ \pm \sqrt{k^2+g_+^2}}{k} \\ e^{i \phi_k } \end{pmatrix},\label{eqn.states1}\\
 & u_{2/4}(\vec k) =\frac{1}{\sqrt{2 +2 \frac{\(g_- \mp \sqrt{k^2+g_-^2}\)^2}{k^2}}}\begin{pmatrix}
 -e^{-i \phi_k } \\\frac{g_-\mp \sqrt{k^2+g_-^2}}{k} \\ -\frac{g_- \mp\sqrt{k^2+g_-^2}}{k} \\ e^{i \phi_k }\end{pmatrix}.\label{eqn.states2}
\end{align}

\section{\large Evolution of bands from CFL$_{2\pi}$ to CFL$_{0}$}
\begin{figure}[h]
 \includegraphics[width=\columnwidth]{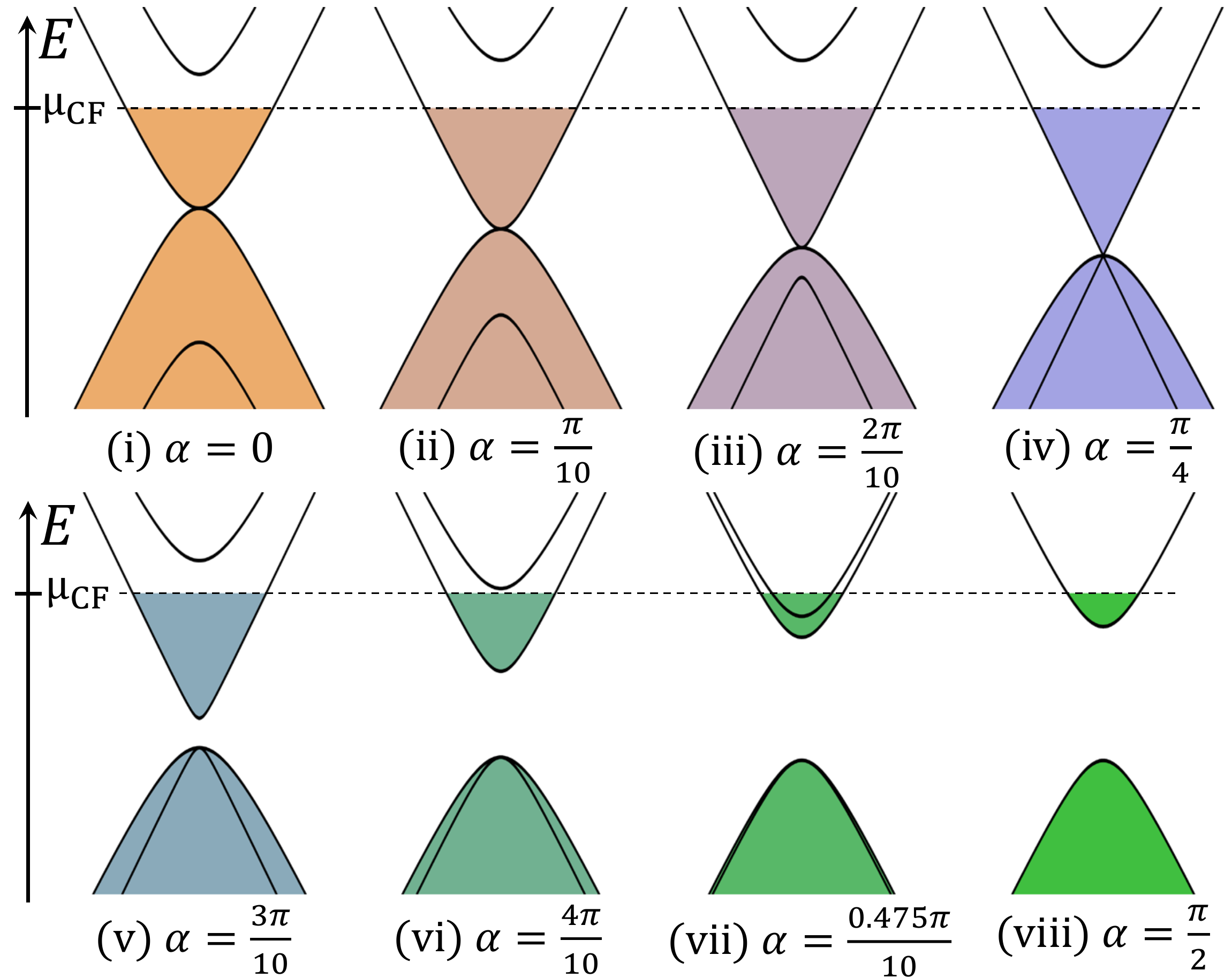}
 \caption{Composite-fermion band structure shown for parameters $g=\cos \alpha$, $\Delta=\sin \alpha$.}
 \label{fig.evol}
\end{figure}

Figure~\ref{fig.evol} shows the composite-fermion band structure for $g = \cos \alpha$, $\Delta = \sin \alpha$ over a range of $\alpha$. (i)-(iii) As $\alpha$ increases from zero, the curvatures of the positive and negative energy bands that meet at the quadratic band touching become unequal. (iv) The transition between CFL$_{2\pi}$ and CFL$_{0}$ occurs at $|\Delta|=|g|$ where three bands meet at one point. (v-viii) For $|\Delta|>|g|$ the spectrum is gapped, and the positive (negative) energy bands become degenerate at $g=0$ . 

\section{\large Regarding $2\pi$ Berry flux}
One may be tempted to argue that $2\pi$ Berry phases are not meaningfully distinct from zero Berry phases. Indeed, in a rotationally symmetric system the Berry phase may be computed as
\begin{align}
\gamma_\text{Berry} = -i \int_{0}^{2\pi} d \theta \langle u | \partial_\theta | u \rangle,\label{eq.berry}
\end{align}
and a gauge transformation 
\begin{align}| u \rangle \rightarrow e^{ i n \theta }| u \rangle \label{eq.gauge}
\end{align}
 changes $\gamma_\text{Berry}$ by $2\pi n $. In this section we explain the sharp distinction between CFL$_{2\pi}$ and CFL$_{0}$ in the presence of inversion symmetry.

First, we note that the above ambiguity only arises in the case of degeneracies; non-degenerate bands always feature a well-defined (smooth) local Berry curvature which can be integrated to a well-defined finite value for $\gamma_\text{Berry} \in \mathbb{R}$ without any compactification. For this reason we refer to $\gamma_\text{Berry}$ here and in the main text as `Berry flux'. In Eq.~\eqref{eq.berry}, the gauge transformation, Eq.~\eqref{eq.gauge}, is singular at the origin and is thus not meaningful for a non-degenerate band. A useful, physical way to resolve this question in the present context is thus to infinitesimally break particle-hole symmetry, while keeping inversion symmetry intact. This procedure results in $\pm 2\pi$ Berry flux in the partially filled positive energy band in the CFL$_{2\pi}$ regime and zero in the CFL$_{0}$ regime; cf.~Eq.~\eqref{eqnberryphases} in the main text. 

Alternatively, one may sharply distinguish between CFL$_{2\pi}$ and CFL$_{0}$ band structures via a \emph{pseudospin} $\hat n$ with
\begin{align}
\hat n = 
\begin{pmatrix}
\hat h_{xx}-\hat h_{yy}\\\hat h_{xy}+\hat h_{yx}
\end{pmatrix},
\end{align}
where $\hat{h}_{\alpha\beta}$ are defined in Eq.~\eqref{eqn.halphabeta}. Evaluating $\hat n$ for the eigenstates $|u_j(\vec k)\rangle$ of the ${\cal C}$- and ${\cal R}(\pi/2)$-invariant Hamiltonian, Eq.~\eqref{eqn.rotandph}, we find
\begin{align}
\langle u_j(\vec k) | \hat n |u_j(\vec k) \rangle = f_{j}\(\vec k\) 
\begin{pmatrix}
\cos 2 \phi_k\\ \sin 2 \phi_k   
\end{pmatrix}.
\end{align}
In the CFL$_{2\pi}$ regime, $g>|\Delta|$, 
\begin{align}
 \lim \limits_{\vec k \rightarrow 0} (f_1,f_2,f_3,f_4) = (0,-1,1,0),
\end{align}
while in the CFL$_{0}$ regime, $\Delta>g>0$, 
\begin{align}
 \lim \limits_{\vec k \rightarrow 0} (f_1,f_2,f_3,f_4) = (0,0,1,-1).
\end{align}
Consequently, the partially filled band, $j=2$, features a winding of the pseudospin in CFL$_{2\pi}$ that is absent in CFL$_{0}$.

\section{\large Projection onto quadratically touching bands}
To make the nature of CFL$_{2\pi}$ more transparent, it is convenient to focus on the two bands that touch quadratically. Writing $\Psi_4^T = (\psi_1,\psi_2,\psi_3,\psi_4)$, the states created by $\psi_1$ and $\psi_4$ form a degenerate subspace at $\vec k=0$. The corresponding states at small but non-zero $|\vec k| \ll |g| -|\Delta|$ are created by
\begin{align}
&\psi_+ = \psi_1+ \frac{ k e^{i \phi_k}}{2 g_+ g_-}\(\Delta \psi_2 - g \psi_4\)\\
&\psi_- = -\psi_4+ \frac{ k e^{i \phi_k}}{2 g_+ g_-}\(g \psi_2-\Delta \psi_4\).
\end{align}
Projecting the Hamiltonian of Eq.~\eqref{eqn.rotandph} onto the two-dimensional subspace spanned by $\psi_\pm$ yields
\begin{align}
P h_{\cal R}(\vec k) P = \frac{1}{2 g_+ g_-}\left(
\begin{array}{cc}
 k^2 \Delta & e^{-2 i \phi_k } g k^2 \\
 e^{2 i \phi_k } g k^2 & k^2 \Delta \\
\end{array}
\right).
\end{align}
The PH-breaking mass term $\sigma^z$ projects as
\begin{align}
&P \sigma^z P = \begin{pmatrix}
1 & 	0 \\
0& -1 
\end{pmatrix}+ \mathcal{O}(k^2).
\end{align}
The fields $\psi_+,\psi_-$ transform under PH symmetry and rotations as
\begin{align}
&{\cal C} \psi_{\pm}{\cal C}^{-1} = - \psi_{\mp},\\
&{\cal R}(\Phi) \psi_{\pm}{\cal R}^{-1}(\Phi) = e^{\pm i \Phi} \psi_{\pm},
\end{align}
which for $\Psi_2^T=(\psi_+,\psi_-)$ gives Eq.~\eqref{psi2transform}.
\section{\large Berry Curvature induced in CFL$_{0}$ by PH symmetry breaking}
To estimate the Berry curvature induced in CFL$_{0}$ by weak breaking of PH symmetry, we consider the limit $\Delta \gg g >0$ and focus on the two positive energy bands with wave functions $u_{1,2}(\vec k)$ specified in Eqs.~\eqref{eqn.states1} and \eqref{eqn.states2}. Weak breaking of particle-hole symmetry $\sim m \sigma^z\tau^0$ with $m \ll g,\Delta$ modifies the wave-functions as 
\begin{align}
\tilde u_{1/2}(\vec k)= u_{1/2}(\vec k)\pm \frac{m}{4g}u_{2/1}(\vec k).
\end{align}
Using these to compute the Berry-flux enclosed in a Fermi-surface of radius $K_F \ll \Delta$ one finds
\begin{align}
\gamma_\text{Berry} =-2 \pi \frac{m}{g} \frac{K_F^2}{8 \Delta^2}
\end{align}
where we expanded to leading order in $K_F$ [cf.~Eq.~\eqref{eqnberryphases} from the main text]. 
\end{document}